\documentclass[11pt]{article}
\usepackage[dvips]{graphicx}  %  XXX archive  version
\begin{document}

\title{\LARGE \bf Analysis of Data from a Quantum Gravity Experiment }  
\author{{Reginald T.
Cahill\footnote{{\bf Process Physics:}
http://www.socpes.flinders.edu.au/people/rcahill/processphysics.html}}\\
 {School of Chemistry, Physics and Earth Sciences}\\
{ Flinders University }\\ 
{ GPO Box 2100, Adelaide 5001, Australia }\\
{(Reg.Cahill@flinders.edu.au)}}

\date{July 1, 2002}
\maketitle

\begin{abstract}
  A new information-theoretic modelling of reality  has
given rise to a quantum-foam description of space, relative to which 
absolute motion is meaningful. In a previous paper (Cahill and Kitto) it was shown that
in this new physics Michelson interferometers show absolute motion effects when operated in
dielectric mode,  as indeed all such experiments had indicated, and  re-analysis of the
experimental data showed that the measured speeds were all in agreement
with the COBE CBR dipole-fit speed of $365\pm18$km/s.  Here  the new physics is applied,
using a different type of analysis, to the extensive data from dielectric-mode
interferometer observations by Miller (1933).  Here the speed of in-flow of the quantum foam
towards the Sun is determined from Miller's data to be $47 \pm 6$km/s, compared to  the theoretical
value of $42$km/s. This observed in-flow is a signature of a quantum gravity effect in the new
physics.
\end{abstract}

\newpage

A new information-theoretic modelling of reality \cite{RC02,RC01}  has
given rise to a quantum-foam description of space, relative to which 
absolute motion is meaningful. In Ref.\cite{CK} it was shown that
in this new physics Michelson interferometers \cite{Mich} show absolute motion effects when
operated in dielectric mode,  as indeed all such experiments had in fact indicated, and 
re-analysis of the experimental data showed that the measured speeds were all in agreement with each
other and together in agreement with the COBE CBR dipole-fit speed of $365\pm18$km/s \cite{Smoot} using the
M\'{u}nera \cite{Munera} re-analysis of interferometer data. The new physics is here further
tested against experiment by re-analysing the extensive dielectric-mode interferometer data of Miller
\cite{Miller2} and extracting a quantum-gravity effect. These results amount to a dramatic development in
fundamental physics.

Although  the theory and experiment together indicate that absolute motion is an aspect of
reality one must hasten to note that this theory also implies that the Einstein Special and
General Theory of Relativity formalism remains essentially intact. In \cite{RC02} it was
shown that this formalism arises from the quantum-foam physics, but that the quantum-foam
system  amounts to a physically real foliation of the spacetime construct.  Despite this
there are some phenomena which are outside the Einstein  formalism, namely the detection of
absolute motion. We see here the emergence of a new theoretical system which subsumes 
the older theory and covers new phenomena, in particular it unifies  gravity and the
quantum phenomena. 

As shown in \cite{CK} and reviewed here the new physics provides a different account of the Michelson
interferometers. The main outcome is that the time difference for light travelling via the
orthogonal arms, when one arm is parallel to the direction of motion, is now given by
\begin{equation}\label{eqn:QG1}
\Delta t=k^2\frac{Lv^2}{c^3}.
\end{equation}
Here $v$ is the magnitude of the absolute velocity ${\bf v}$ of the interferometer through the
quantum-foam, projected onto the plane of the interferometer.  And $k=\sqrt{n(n^2-1)}$  where $n$ is the
refractive index of the medium through which the light in the interferometer passes,  $L$ is the length of each
arm and
$c$ is the speed of light relative to the quantum foam. This expression follows from both the
Fitzgerald-Lorentz contraction effect and that the speed of light through the dielectric is
$V=c/n$, ignoring any Fresnel dragging effects. This is one of the  aspects of the quantum foam
physics that distinguishes it from the Einstein formalism and is discussed in \cite{CK}. The time difference
$\Delta t$ is revealed by the fringe shifts on rotating the interferometer. In Newtonian physics
 $k=\sqrt{n^3}$ \cite{CK,Mich}, while in Einsteinian physics $k=0$ expressing the fundamental
assumption that absolute motion is not measurable and indeed has no meaning. So the experimentally
determined value of
$k$ is a key test of fundamental physics. 
  
In deriving (1) in the new physics it is essential to note that space is a quantum-foam system
\cite{RC02,RC01}  which exhibits various subtle features. In particular it exhibits real dynamical
effects on clocks and rods. In this physics the speed of light is only $c$  relative to the
quantum-foam, but to observers moving with respect to this quantum-foam the speed appears to be still $c$, but
only because their clocks and rods are affected by the quantum-foam. As shown in \cite{RC02} such observers
will find that observations of distant events will be described by the Einstein spacetime formalism, but only
if they restrict measurements to those achieved by using clocks, rods and light pulses.   It is simplest in
the new physics to work in the quantum-foam frame of reference.  If there is a dielectric present at rest in
this frame, such as air, then the speed of light in this frame is
$V=c/n$. If the dielectric is moving with respect to the quantum foam, as in an interferometer attached to the
Earth, then the speed of light relative to the quantum-foam is still $V=c/n$ up to corrections due to Fresnel
drag.  But this dragging is a  very small effect and is  not required in the present analysis.  Hence  this
new physics requires a different method of analysis from that of the Einstein physics. With these cautions we
now describe the operation of a Michelson interferometer in this new physics, and show that it makes
predictions different to that of the Einstein physics.    Of course experimental evidence is the final
arbiter in this conflict of theories.  

\vspace{-3mm}
\begin{figure}[h]
\hspace{25mm}\includegraphics[scale=0.6]{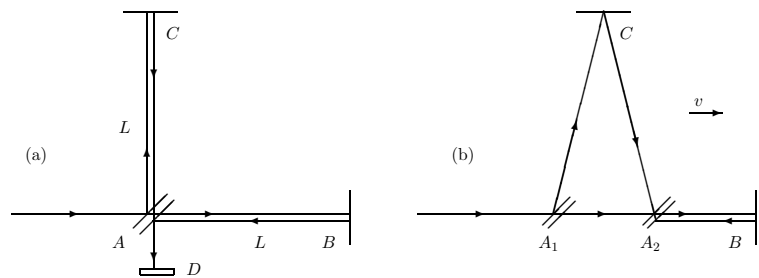}
\vspace{27mm}
\caption{\small{Schematic diagrams of the
Michelson Interferometer, with
beamsplitter/mirror at $A$ and mirrors at $B$ and
$C$, on equal length arms when parallel,
from $A$. $D$ is a quantum detector (not drawn in (b)) that causes localisation 
of the photon state by a collapse process. In (a)
the interferometer is at rest in space. In (b) the
interferometer is moving with speed $v$
relative to space in the direction
indicated. Interference
fringes are observed at the quantum detector $D$. 
  If the interferometer is
rotated in the plane  through $90^o$, the
roles of arms
$AC$ and $AB$ are interchanged, and during
the rotation shifts of the fringes are seen
in the case of absolute motion, but only if the apparatus operates in a dielectric.  By counting
fringe changes the speed $v$ may be
determined.}}
\end{figure}

As shown in Fig.1 the  beamsplitter/mirror $A$ sends a photon $\psi(t)$ into a superposition
$\psi(t)=\psi_1(t)+\psi_2(t)$, with each component travelling in different arms of the interferometer, until
they are recombined in the quantum detector (D) which results in a localisation process, and one spot in the
detector is produced.  Repeating with many photons reveals that the interference between $\psi_1$ and $\psi_2$
at the detector results in fringes.   To simplify the analysis here assume that the two arms are constructed to
have the same lengths   when they are physically parallel to each other. The Fitzgerald-Lorentz effect in the
new physics  is that the arm $AB$  parallel to the direction of motion is shortened to
\begin{equation}\label{eq:MM2}
L_{\parallel}=L\sqrt{1-\frac{v^2}{c^2}}
\end{equation}
by motion through space at speed $v$.
Following Fig.(1) we consider the case when  the apparatus is moving at speed $v$ through space,
and  that the photon states  travel at speed $V=c/n$ relative to the quantum-foam which is space, where $n$ is the
refractive index of the gas and
$c$ is the speed of light, in vacuum,  relative to the space.  Let the time taken for $\psi_1$ to travel from
$A\rightarrow B$ be
$t_{AB}$ and that from $B\rightarrow A$ be $t_{BA}$. 
In moving from the beamsplitter at $A$ to $B$, the photon state $\psi_1$ must travel an extra  distance 
because
 the mirror $B$ travels a distance $vt_{AB}$ in this time, thus the total distance that must be traversed  is
\begin{equation}\label{eq:MM3}
Vt_{AB}=L_{\parallel}+vt_{AB}.
\end{equation}
Similarly on returning from $B$ to $A$ the photon state $\psi_1$ must travel the distance
\begin{equation}\label{eq:MM4}
Vt_{BA}=L_{\parallel}-vt_{BA}.
\end{equation}
Hence the total time $t_{ABA}$ taken for $\psi_1$  to travel from $A\rightarrow B \rightarrow A$ is given
by
\begin{eqnarray}\label{eq:MM5}
t_{ABA}=t_{AB}+t_{BA}&=&\frac{L_{\parallel}}{V-v}+\frac{L_{\parallel}}{V+v}\\
&=&\frac{L_{\parallel}(V+v)+L_{\parallel}(V-v)}{V^2-v^2}\\
&=&\frac{2LV\sqrt{1-\displaystyle\frac{v^2}{c^2}}}{V^2-v^2}.
\end{eqnarray}
Now let the time taken for the photon state $\psi_2$ to travel from $A\rightarrow C$ be $t_{AC}$,
but in that time the apparatus travels a distance $vt_{AC}$.  Pythagoras' theorem then gives
\begin{equation}\label{eq:MM8}
\left(Vt_{AC}\right)^2=L^2+\left(vt_{AC}\right)^2
\end{equation}
which gives
\begin{equation}\label{eq:MM9}
t_{AC}=\frac{L}{\sqrt{V^2-v^2}},
\end{equation}
and including the return trip $C\rightarrow A,  t_{CA}=t_{AC}, t_{ACA}=t_{AC}+t_{CA}$ results in 
\begin{equation}\label{eq:MM10}
t_{ACA}=\frac{2L}{\sqrt{V^2-v^2}},
\end{equation}
giving finally for the time difference for the two arms
\begin{equation}\label{eq:MM11}
 \Delta t= \frac{2LV\sqrt{1-\displaystyle\frac{v^2}{c^2}}}{V^2-v^2}-\frac{2L}{\sqrt{V^2-v^2}}.
\end{equation}
Now trivially $\Delta t =0$  if $v=0$, but  also $\Delta t =0$ when $v\neq 0$ but only if $V=c$.  This then 
  would  result in a null result on rotating the apparatus.  Hence the null result of the Michelson-Morley
apparatus in the new physics is only for the special case of photons travelling in vacuum for which $V=c$.   
However if the Michelson-Morley apparatus is immersed, for example, in a gas then
$V<c$ and a non-null effect is expected on rotating the apparatus, since now  $\Delta t \neq 0$.  It is
essential then in analysing data to correct for this refractive index effect.  Putting
$V=c/n$ in (11) we find for $v << V$  that 
\begin{equation}\label{eq:MM12}
\Delta t= Ln(n^2-1)\frac{v^2}{c^3}+\mbox{O}(v^4),
\end{equation}
that is $k=\sqrt{n(n^2-1)}$,  which gives $k=0$ for vacuum experiments ($n=1$). 

 However if the data from dielectric mode interferometers is (incorrectly)  analysed not using the
Fitzgerald-Lorentz contraction (2), then, as done in the old analyses,   the estimated Newtonian-physics time
difference is 
\begin{equation}\label{eq:MM13}
\Delta t = \frac{2LV}{V^2-v^2}-\frac{2L}{\sqrt{V^2-v^2}},
\end{equation}
which again for $v << V$  gives 
\begin{equation}\label{eq:MM14}
\Delta t = Ln^3\frac{v^2}{c^3}+\mbox{O}(v^4),
\end{equation}
that is $k=\sqrt{n^3}$. The value of $\Delta t$ is deduced from analysing the fringe shifts, and then    the speed
$v_{M}$ (in previous Michelson-Morley type analyses) has been extracted  using (\ref{eq:MM14}), instead of the
correct form (\ref{eq:MM12}). 
$\Delta t$
 is typically of order $10^{-15}s$  in gas-mode interferometers, corresponding to a fractional fringe
shift.   However it is very easy to correct for this oversight.  From (\ref{eq:MM12}) and (\ref{eq:MM14}) we
obtain, for the corrected absolute speed $v$ through space, and for $n \approx 1^+$, 
\begin{equation}\label{eq:MM15}
v=\frac{v_{M}}{\sqrt{n^2-1}}.
\end{equation}
  
Of the early interferometer experiments  Michelson and Morley
\cite{MM} and Miller \cite{Miller2}  operated in air ($n=1.00029$), while that of  Illingworth
\cite{Illingworth} used Helium ($n=1.000035$). We expect then that for air interferometers
$k_{air}^2=0.00058$ (i.e.  $k_{air}=0.0241$) and for Helium $k_{He}^2=0.00007$, which explains why these
experiments reported very small but nevertheless non-null and so significant effects.
 All non-vacuum experiments gave
$k>0$, that is, a non-null effect. All vacuum ($n=1$) interferometer experiments, having $k=0$, give null
effects as expected,  but such experiments cannot distinguish between the new physics and the Einstein physics,
only dielectric-mode interferometers can do that. The notion that the Michelson-Morley experiment gave
a null effect is a common misunderstanding that has dominated  physics for more than a century. By
``null effect'' they meant that the effect was much smaller than expected, and the cause for this is
only now apparent from the above. When the air and Helium interferometer data were re-analysed using
the appropriate $k$ values in \cite{CK} they gave consistent values which were also in agreement with
the CBR speed.  So these early interferometer experiments did indeed reveal absolute motion, and
demonstrated that $k \neq 0$ in these experiments.

Here the data from the exquisite and extensive  air-interferometer  experiments with $L=64$m
(obtained by multiple reflections along the arms)  by  Miller at Mt.Wilson (Latitude $+34^0 13'$)
beginning in 1921
\cite{Miller2} are  re-analysed using not only the quantum-foam effect in (\ref{eqn:QG1}), but also a
quantum-gravity effect predicted by the new physics.   It turns out that Miller actually performed the
first quantum gravity experiment.   Miller reported in
\cite{Miller2}  particular  observations  over four days in 1925/26 recording the time variation of
the projection of the velocity  ${\bf v}$ onto the interferometer throughout each of these days.   His
data is shown in Fig.2 for  the azimuths  and $v_M$ speeds obtained from fringe shifts. The azimuth is
the angle  measured from the  local meridian  indicating the direction of the projected velocity $\bf
v$, and the $v_M$  speed is  defined as that from (\ref{eqn:QG1}) with $k=1$ using the measured $\Delta t$, so
that $v=v_M/k$.  

Miller's idea was that ${\bf v}$ should have two components: (i) a cosmic velocity of the solar
system through space, and (ii) the orbital velocity of the Earth about the Sun through space. Over a year this
vector sum  would result in a changing ${\bf v}$, as was in fact observed, and is obvious in Fig.2.  Further,
since the  orbital speed was known, Miller was able to extract from the data the magnitude and
direction of
${\bf v}$ as the orbital speed offered an absolute scale. We shall use  $\overline{k}$ as the
$k$ value obtained by this approach. Miller was led to the conclusion that for reasons unknown  the
interferometer did not indicate true values of
$v$, and for this reason  he introduced the parameter $\overline{k}$.   Miller noted, in fact, that
$\overline{k}^2<<1$. Fitting  the  data using $v_{tangent}=30$km/s we obtain  $\overline{k}=0.044\pm
0.005$ (Miller found $\overline{k}=0.046$) and $v=210$km/s and a direction (see later). However  that
$\overline{k} > k_{air}$ tells us that another velocity component has been overlooked.   Miller only knew of the
tangential or orbital speed of the Earth, whereas the new physics predicts that as-well there is a
quantum-gravity radial in-flow ${\bf v}_{in}$ of the quantum foam, so  
\begin{equation}\label{eqn:QG2}
{\bf v}= {\bf v}_{cosmic} +{\bf v }_{tangent} -{\bf v}_{in}.
\end{equation}
\newpage
.
%\vspace{0mm}
\begin{figure}[ht]
\hspace{20mm}\includegraphics[scale=1.5]{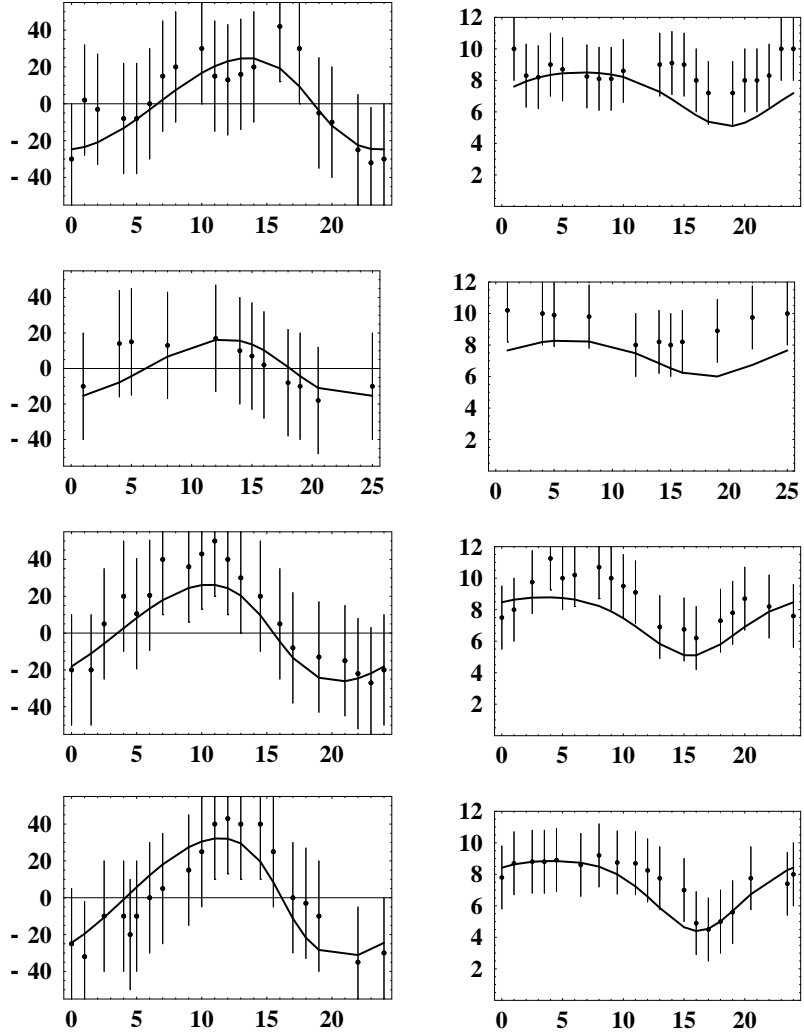}
\vspace{0mm}
\caption{\small Azimuths (degrees)  (left column) and $v_M$ speeds (km/s) (right column) measured by
Miller  on February 8, 1926 (top row), April 1, 1925, August 1, 1925 and September 15, 1925 (bottom
row) plotted against sidereal time (hrs).  Error bars are indicative only. Curve shows quantum-foam
theory predictions for case of CBR cosmic speed of $365$km/s, orbital velocity of $30$km/s,  quantum gravity
in-flow of
$42$km/s, the $k_{air}$ value and Miller's direction for ${\bf v}_{cosmic}$.} 
\end{figure}  

 We can combine the dielectric effect and the Earth-velocity effect to extract from Miller's data the
speed of this in-flow component. For circular orbits the in-flow and tangential speeds are orthogonal,
 so the actual velocity of the Earth through the quantum foam is given by (\ref{eqn:QG2})
 whereas as  Miller of course did not include the  $-{\bf v}_{in}$ component.  From this we easily find that it
is $v_R=\sqrt{v_{in}^2+v^2_{tangent}}$ that sets the scale and not $v_{tangent}$, and so we obtain that the value
of $v_{in}$ implied by  $\overline{k}>k_{air}$  is given by
\begin{equation}\label{eqn:QG3}
v_{in}=v_{tangent}\sqrt{\displaystyle{ \frac{\overline{k}^2}{k_{air}^2}-1 }}
\end{equation} 
 Using the above
$\overline{k}$  value and the value of $k_{air}$  we obtain $v_{in}=47 \pm 6$km/s.
The new physics which unifies gravity and the quantum predicts  the in-flow speed to be, at distance
$R$ from the Sun,
\begin{equation}\label{eqn:QG4}
v_{in}=\sqrt{\displaystyle{\frac{2GM}{R}}},\end{equation}    
where $M$ is the mass of the Sun, $R$ is the distance of the Earth from the
Sun, and $G$ is Newton's gravitational constant. $G$ is essentially a measure of the rate at
which matter effecctively `dissipates' the quantum-foam.  For circular orbits the tangential orbital speed
is given by
\begin{equation}\label{eqn:QG5}
v_{tangent}=\sqrt{\displaystyle{\frac{GM}{R}}},\end{equation}
giving $v_{tangent}=30$km/s, 
and so $v_{in} =\sqrt{2}v_{tangent}=42$km/s. Hence the value for $v_{in}$ from the Miller data is 
consistent with the theoretical value. Since it is $v_R=\sqrt{3}v_{tangent}$ and not $v_{tangent}$ that sets the
scale we must re-scale Miller's value for $v$ to be $\sqrt{3}\times 210=364$km/s, which now compares favourably
with the COBE CBR speed.

The gravitational acceleration arises from inhomogeneities in the flow
and is given by \newline${\bf g}=({\bf v}_{in}.{\bf  \nabla }){\bf v}_{in}$ in this quantum-foam flow physics
\cite{RC02}. So the Miller experiment actually amounted to the first quantum gravity experiment, and the ability
of dielectric-mode interferometers to measure absolute motion made this possible. New dielectric-mode
interferometers are being used at Flinders University to measure the quantum-foam radial in-flow speed
associated with the Earth's gravity. This component did not contribute to Miller's data as it is perpendicular
to the plane of the interferometer.

Miller was also able to extract the direction of ${\bf v}_{cosmic}$ from analysis of
his observational data using the daily and seasonal time-dependencies apparent in Fig.2
 and obtained  a right ascension
and declination of $(\alpha,\delta)=(17^h, +70^0)$.  In Fig.2  are shown the
quantum-foam physics `predictions' for  the $v_M$ speeds and azimuths  using the projection of ${\bf v}$ in
(\ref{eqn:QG2})  onto the plane of the Miller interferometer at Mt.Wilson,
using $v_{cosmic}= v_{CBR}$  is $365$km/s, the orbital speed
is $30$km/s and the in-flow speed is $42$km/s,   the  $k_{air}$ value, and  the Miller direction for
${\bf v}_{cosmic}$. This  illustrates that the new physics agrees with Miller's 1925/1926 observational data,
both for absolute values and for time variation over each day and seasonally.  The Miller data used here was
obtained by inspection of hand-drawn figures published in \cite{Miller2} and are not very accurate, and so a
complete re-analysis of the data is not possible. The error bars are only indicative and should not be taken as
Miller's values.  

While the extracted cosmic speed from Miller's data here, and from the Michelson-Morley and
Illingworth data in \cite{CK}, are in excellent agreement with the CBR speed  the direction
found above is in complete disagreement with the COBE determined direction \cite{Smoot} which is
 $(\alpha,\delta)=(11.2^h\pm 0.2^h, -7^0 \pm 2^0)$. This COBE direction gives
azimuth and $v_M$ speed interferometer predictions that are totally inconsistent with the Miller data
in Fig.2. However the two directions are at $90^0$ to each other. This suggests that
this disagreement might be  caused by a definitional problem with the right ascension and
declination  in the COBE satellite mission. At least this is a prediction of the present
work.

It is interesting to return to the Michelson-Morley experimental data of 1887 which  since
then, with rare exceptions, has been claimed to have given a null result, and so
supporting the Einstein assumption that absolute motion has no meaning.  In fact
Michelson-Morley  reported non-null effects, but much smaller than they expected.  They made
observations of thirty-six complete turns  using a $L=1.1$ meter length air-interferometer in
Cleveland, Latitude $41^0 30'$N, with six turns at  $12\!:\!00$ hrs ($7\!\!:\!\!00$ hrs ST) on each day of July 8,
9 and 11, 1887  and similarly at $18\!:\!00$ hrs ($13\!\!:\!\!00$ hrs ST) on July 8, 9 and 12, 1887.  Their
results  after averaging over  the noon sessions,  and also averaging the evening sessions, and also
averaging the first half-turn and the last half-turn data, are shown in Fig.3 (the data is available from
\cite{Miller2}).

The fringe shifts were extremely small but within their observational capabilities.  The fringe shifts
actually correspond to a  maximum $v_M$ speed of some $8.3$km/s, being slighlty lower than Miller's speeds in Fig.2
due to the higher latitude of Cleveland compared to Mt.Wilson.  Also plotted in Fig.3  are the quantum-foam physics
predictions using in (\ref{eqn:QG2}) the  CBR speed of $365$km/s, the orbital speed of $30$km/s, the quantum 
gravity in-flow of
$42$km/s, the Miller direction and the value of
$k_{air}$, for the time, date and location of the Michelson-Morley experiment. The dependence on  rotation angle
$\theta$ in Fig.3 is $sin(2\theta)$. We see that they observed essentially what the new physics predicts. 

\vspace{2mm}
\begin{figure}[ht]
\hspace{0mm}\includegraphics[scale=1.1]{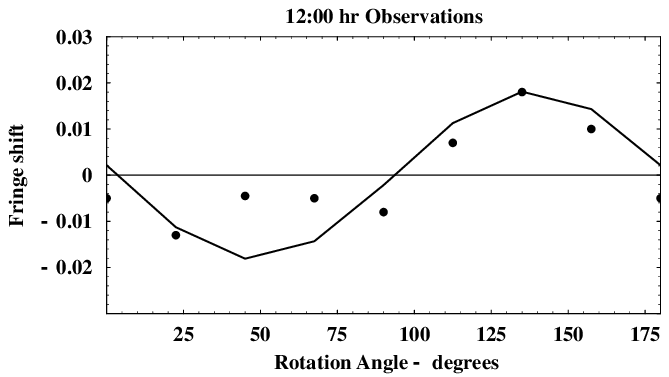}
\hspace{2mm}\includegraphics[scale=1.1]{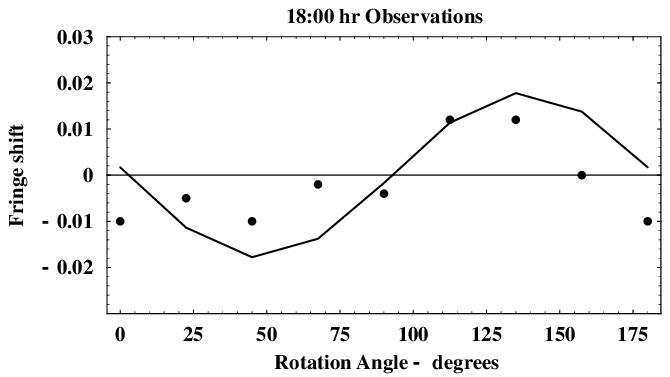}
\vspace{-3mm}
\caption{\small  Data shows the 1887 Michelson-Morley averaged  fringe shifts for $12\!\!:\!\!00$ hrs 
 on July 8, 9 and 11 (LH plot)  and $18\!\!:\!\!00$ hrs  on July 8, 9 and 12 (RH plot),
as  interferometer was rotated through 360 degree, and is average of the two half-turns. Curve 
shows quantum-foam theory prediction for case of CBR cosmic speed $365$km/s, orbit speed of $30$km/s, 
quantum gravity in-flow of $42$km/s, the $k_{air}$ value and Miller's direction for ${\bf v}_{cosmic}$.} 
\end{figure} 

So their results were never null, and can now be fully comprehended.
They  in-fact performed the first quantum-foam experiment, as revealed by the dielectric effect. Their expected
fringe shifts  were based on using the Newtonian value of $k=1$ and on $v$ being atleast $30$km/s, due to the
Earth's orbital motion, and so predicting fringe shifts 10 times larger than actually seen (the true value for 
$v^2$ in (\ref{eqn:QG1}) is some $10^2$  larger but  the dielectric effect gives a reduction of approximately 
$1/1000$).  Of course that Michelson and Morley saw any effect is solely due to the presence of the air in their
interferometer, an effect that neither they nor later generations of physicists ever noticed.  Vacuum
interferometer experiments of the same era by Joos
\cite{vacuum} gave  
$v_M<1$km/s, and are consistent with a null effect, as also predicted by the quantum-foam physics. 
Of course if Michelson and Morley had used glass rods in their interferometer  they would have seen effects
more than 1000 times larger, and the history of physics over the last 100 years would have been totally
different.  

The experimental results analysed herein and in \cite{CK} show that absolute motion is detectable.  
This is motion with respect to  a quantum-foam system that is space. As well quantum matter
effectively acts as a sink for the quantum-foam, and the flow of that quantum-foam towards the Sun has
been confirmed in Miller's data. So that experiment was the first quantum gravity experiment. These
results are in conflict with the fundamental assumption by Einstein that absolute motion has no
meaning and so cannot be measured.  Vacuum interferometer experiments do give null results, for
example see  \cite{vacuum,  KT, BH, Muller}, but they only check the Lorentz contraction effect, and this is
common to both theories. So they are unable to distinguish the new physics from the Einstein physics.  As well
that the interferometer experiments and their results fall into two classes, namely vacuum and dielectric has
gone unnoticed. The non-null results from dielectric-mode interferometers have always been rejected on the 
grounds that they would be in conflict with the many successes of the Special and General Theory of
Relativity.   However this is not strictly so, and it turns out that these successes survive in the new
physics, which actually subsumes the Einstein formalism, even though the absolute motion effect is not in the
Einstein physics. Einstein essentially arrived at a  valid formalism from a wrong assumption.  The new more
encompassing physics allows the determination of a physically real foliation of the spacetime construct (the
Panlev\'{e}-Gullstrand foliation)  and so it actually breaks the diffeomorphism symmetry of General Relativity.
The results here and in \cite{CK} demonstrate that the new physics  is an experimentally
tested  unification of gravity and the quantum.

This paper is dedicated to the memory of Dayton C. Miller of the Case School of Applied Science,
Cleveland, Ohio. The author thanks Warren Lawrance for on-going discussions of new-generation
dielectric-mode interferometer experiments.

\end{document}